\documentclass[preprint,preprintnumbers,showpacs,aps,amssymb,showkeys]{revtex4}

\usepackage{graphicx}
\usepackage{bm}
\usepackage{amsmath}

\def\calL{{\cal L}}
\def\calO{{\cal O}}
\def\calP{{\cal P}}

\def\bbar{{\bar b}}
\def\hbar{{\bar h}}
\def\nbar{{\bar n}}
\def\qbar{{\bar q}}

\def\ubar{{\bar u}}
\def\Lmdbar{{\bar \Lambda}}

\def\vslash{v\hspace{-1.8mm}/}
\def\nslash{n\hspace{-2.2mm}/}
\def\nbarslash{\nbar\hspace{-2.2mm}/}

\def\Bslash{B\hspace{-2.7mm}/}
\def\Dslash{D\hspace{-2.7mm}/}

\begin{document}

\title{Light-cone sum rules for the heavy-to-light transition in the effective
theory}

\author{Jong-Phil Lee}
\email{jongphil@korea.ac.kr}
\affiliation{Department of Physics, Korea University, Seoul, 136-701, Korea}

\begin{abstract}
Heavy-to-light weak form factor is calculated using the light-cone sum rule (LCSR)
in the framework of soft-collinear effective theory (SCET).
There are spin-symmetric and spin-nonsymmetric contributions.
Leading order spin-symmetric contribution corresponds to the ''soft overlap''
where some of the partons carry very small momentum.
The next-to-leading order spin-symmetric and spin-nonsymmetric parts are
characterized by a collinear gluon exchange with the spectator quark.
We reproduce the full theory LCSR results and give comments on recent LCSR in
SCET.
\end{abstract}


\maketitle
\section{Introduction}
Heavy-to-light decay processes become more and more important nowadays with
copious data from the $B$ factories.
Especially, $B\to \pi(\rho)$ decays provide information about the poorly known
Cabibbo-Kobayashi-Maskawa (CKM) element $V_{ub}$.
In a theoretical viewpoint, heavy-to-light transition shows interesting
kinematical configurations which provoke new effective theories of strong
interaction.
Soft-collinear effective theory (SCET) is a useful framework to deal with light
and energetic particles \cite{SCET}.
\par
But nonperturbative quantities are always involved in the analysis.
The way how to deal with them is crucial and distinctively
differs in various approaches.
Among the nonperturbative calculational methods, light-cone sum rule (LCSR) has 
been very successful
\cite{Balitsky:1989ry,Belyaev:1993wp,Ball:1997rj,Ball:1998kk,Bagan:1997bp,Ball:1998tj,
Ball:2001fp,Ball:2003fq,Ball:2004ye}.
Compared to the traditional sum rule by Shifman, Vainshtein, and Zakharov (SVZ)
\cite{SVZ}, LCSR is more adequate for the heavy-to-light decays
\cite{Ball:1997rj}.
In the SVZ sum rules, nonperturbative effects are encoded by the so called
vacuum condensates.
But it is argued that the condensates exaggerate the end-point behavior of
the final state meson distribution amplitudes (DA) in the heavy-to-light decays
\cite{Ball:1997rj}.
The LCSR is based on an expansion of nonlocal operators in twist whose matrix
elements between vacuum and the meson define the meson distribution amplitudes.
\par
One of the most representative nonperturbative quantity for the heavy-to-light
transition is the $B\to\pi$ form factor, $f_+$.
It was already calculated by LCSR in \cite{Bagan:1997bp}.
A new development of QCD factorization (QCDF) \cite{BBNS} enlarged our 
understanding of various $B$ decays, and the spectator interactions are 
systematically examined in \cite{Beneke:2000wa}.
The advent of SCET enabled us to analyze $f_+$ more profoundly
\cite{Beneke:2002ph,Hill:2002vw,Beneke:2003pa,Lange:2003pk,Bauer:2002aj,Bauer:2004tj}.
\par
What makes the analysis more complicated is the possible existence of the
Feynman mechanism, or soft overlap.
In this situation, some of the partons carry very small momentum compared to
others to make the final state meson.
It is quite controversial how large the soft overlap is, or even how to
{\em define} this soft contribution \cite{Bauer:2004tj,Beneke:2004bn,Bauer:2005wb}.
For example, $\zeta^{B\pi}$ in \cite{Bauer:2004tj} includes only collinear
quarks for the final pion, implicitly neglecting the soft-overlap configuration.
That is the reason why the $\zeta^{B\pi}$ is power-counted from
$\alpha_s(\sqrt{Q\Lambda})$, where $Q$ $(\Lambda)$ is a large (hadronic) scale.
On the other hand, the ''soft form factor'' $\xi_\pi$ in \cite{Beneke:2003pa}
includes a hard-collinear spectator interactions as its $\alpha_s(\sqrt{Q\Lambda})$
corrections which are not momentum asymmetric.
And the authors of \cite{Lange:2003pk} define the ''universal form factor''
$\zeta_M$ in terms of soft-collinear messenger modes to describe the soft
overlap.
A clearer and more definite construction will be needed to remove any confusions.
We will use the terminology of ''spin-symmetric'' and ''spin-nonsymmetric''
contributions to the form factors \cite{Ball:2003bf}.
\par
Recently, there was a try to establish the LCSR in SCET to calculate the
heavy-to-light form factors \cite{DeFazio:2005dx}.
Actually, there have been many efforts to construct the LCSR in the effective
theory, like HQET \cite{Korner:2002ba,Wang:2001mi}.
In a conventional method, one constructs the correlation function typically
with the interpolating current for the initial state and the weak current.
After possible contractions, the remaining fields between the vacuum and the
final state define the distribution amplitudes.
The authors of \cite{DeFazio:2005dx} claim that when the momentum configuration
of the final partons is highly asymmetric, then the light-cone expansion of
the remaining fields is not guaranteed.
They propose an alternative method where the final state pion is described by
the interpolating fields.
But as we will see later, new method for LCSR does not show a fundamental
difference from the conventional one.
Rather, the two methods are {\em equivalent} under proper conditions.
\par
In this paper, a natural extension of the conventional LCSR in the effective
theory is pursued.
In other words, we describe the initial $B$ meson in terms of the usual
interpolating fields.
We reproduce the previous results for the form factors of LCSR.\\
As for the soft form factors, they will be still suffering from the criticism
from \cite{DeFazio:2005dx}.
But we argue that the defects are not fully overcome by a new method also.
Still the essential point is how to combine the momentum-asymmetric partons
into the final state pion.
In spite of this difficulty, we would like to show that LCSR in SCET can be well
established, especially when the collinear gluon is exchanged with the spectator
quark, hoping a deeper understanding about SCET, much like the traditional SVZ
sum rule in the HQET.
\par
The paper is organized as follows.
In the next section, the basic formalism of SCET and LCSR is summarized.
Sum rule calculations are given in Sec.\ III.
We distinguish spin-symmetric and spin-nonsymmetric contributions to the form
factor, and reproduce the well known results.
Discussions and conclusions appear in Sec.\ IV.

\section{Setup}
Effective fields in SCET include collinear quark fields $\xi_{n,p}$,
ultrasoft(usoft) heavy quark fields $h_v$, usoft light quark fields $q_{us}$,
collinear gluon fields $A_{n,q}^\mu$, and usoft gluon fields $A_{us}^\mu$.
These are the relevant degrees of freedom for the usoft-collinear interactions.
Collinear covariant derivatives can be defined as
\begin{equation}
i\nbar\cdot D_c=\bar{\calP}+g\nbar\cdot A_n~,~~~
iD_c^{\perp\mu}=\calP_\perp^\mu+gA_n^{\perp\mu}~,
\end{equation}
where $\calP^\mu$ is the usual label operator \cite{SCET,Pirjol:2002km},
and usoft covariant derivatives are
\begin{equation}
i\nbar\cdot D_{us}=i\nbar\cdot\partial+g\nbar\cdot A_{us}~,~~~
iD_{us}^{\perp\mu}=i\partial_\perp^\mu+gA_{us}^{\perp\mu}~.
\end{equation}
The $n$-components of the derivative appear as \cite{Pirjol:2002km}
\begin{equation}
in\cdot D=in\cdot\partial+gn\cdot A_n+gn\cdot A_{us}~.
\end{equation}
\par
We begin with the SCET Lagrangians for the usoft-collinear interactions
\cite{Pirjol:2002km,Bauer:2003mg} :
\begin{eqnarray}
\calL_{\xi q}^{(1)}&=&\bar{\xi}_n\frac{1}{i\nbar\cdot D_c}ig\Bslash_c^\perp
 Wq_{us}+{\rm h.c.}~,\nonumber\\
\calL_{\xi q}^{(2a)}&=&\bar{\xi}_n\frac{1}{i\nbar\cdot D_c}ig n\cdot M~
 Wq_{us}+{\rm h.c.}~,\nonumber\\
\calL_{\xi q}^{(2b)}&=&\bar{\xi}_n\frac{\nbarslash}{2}i\Dslash_c^\perp
\frac{1}{(i\nbar\cdot D_c)^2}ig\Bslash_c^\perp Wq_{us}+{\rm h.c.}~,
\end{eqnarray}
where the field strength operators are defined as
\begin{eqnarray}
igB_c{^\perp\mu}&=&[i\nbar\cdot D^c, ~iD_c^{\perp\mu}]~,\nonumber\\
ign\cdot M&=&\left[i\nbar\cdot D^c,~in\cdot D\right]~,
\end{eqnarray}
and $W$ is the collinear Wilson line,
\begin{equation}
W=\left[\sum_{\rm perm}\exp\left(-\frac{g}{\bar{\calP}}\nbar\cdot A_{n,q}(x)
\right)\right]~.
\end{equation}
The superscripts on $\calL_{\xi q}$ are the power suppression in
$\lambda=\Lambda_{\rm QCD}/Q$.
\par
Heavy-to-light currents are constructed in SCET by matching from full QCD.
Up to the next-to-leading order (NLO) in $\lambda$ at tree level, we have
\begin{eqnarray}
J^{(0)}&=&\bar{\xi}_n W \Gamma h_v~,\nonumber\\
J^{(1a)}&=&-\bar{\xi}_n\frac{\nbarslash}{2}i\overleftarrow{\Dslash}_c^\perp
           W\frac{1}{\bar{\calP}}\Gamma h_v~,\nonumber\\
J^{(1b)}&=&-\bar{\xi}_n\Gamma\frac{\nslash}{2}i\Dslash_c^\perp W
           \frac{1}{m_b}h_v~.
\end{eqnarray}
\par
The essence of LCSR is to calculate the correlation function
\begin{equation}
\Pi=i\int d^4x~e^{-ip_B\cdot x}~\langle\pi(p)|T[J(0)j_B^\dagger(x)]|0\rangle~.
\label{Pi}
\end{equation}
The heavy-to-light current $J(x)$ and the $B$-meson interpolating field
$j_B^\dagger$ are
\begin{equation}
J(x)=\qbar(x)\Gamma b(x)~,~~~j_B^\dagger=\bbar(x)i\gamma_5 q(x)~.
\end{equation}
Using the effective fields, the correlator $\Pi$ can be written in the hadronic
language as
\begin{equation}
\Pi=\frac{F_B}{\sqrt{m_B}}
 \frac{\langle\pi|J_{\rm eff}|B\rangle}{2m_B-\eta-\omega-i\epsilon}+({\rm res.})~,
\end{equation}
where $\eta\equiv 2v\cdot p$, $\omega\equiv 2v\cdot q$, and
$J_{\rm eff}(x)={\bar\xi}_n(x)\Gamma h_v(x)$.
$B$-meson effective decay constant $F_B$ is defined by
\begin{equation}
\langle B|\hbar_v i\gamma_5 q|0\rangle=\sqrt{m_B}F_B=\frac{m_B^2}{m_b}f_B~.
\end{equation}
By the dispersion relation and quark-hadron duality, the LCSR is established as
\begin{equation}
\frac{F_B}{\sqrt{m_B}}
\frac{\langle\pi|J_{\rm eff}|B\rangle}{2m_B-\eta-\omega-i\epsilon}
=\frac{1}{\pi}\int_{m_b-\Lmdbar}^{s_0}\frac{{\rm Im}\Pi(s,\eta)}{s-\omega}ds~,
\end{equation}
where $\Lmdbar=m_B-m_b$,
and the Borel transformed version is
\begin{equation}
\frac{F_B}{\sqrt{m_B}}\exp\left(-\frac{2m_B-\eta}{T}\right)
\langle\pi(p)|J_{\rm eff}|B\rangle=
\frac{1}{\pi}\int_{m_b-\Lmdbar}^{s_0}~e^{-s/T}~{\rm Im}\Pi(s,\eta)~ds~,
\label{BT}
\end{equation}
where $T$ is the Borel parameter.
Note that we use $\omega=2v\cdot q$ as a dynamical variable rather than the
usual $2v\cdot k=2v\cdot (p_B-m_b v)$, so the lower limit of the dispersion
integral is $m_b-\Lmdbar$ which corresponds to $2v\cdot k=0$ when
$\eta=2v\cdot p=m_B$.
The matrix elements $\langle\pi|J_{\rm eff}|B\rangle$ are proportional to the
effective form factors.
In the next section, we evaluate ${\rm Im}\Pi(s,\eta)$ with the quark-gluon fields
for various $J^{(m)}$ to complete the sum rule.

\section{Sum rules in the effective theory}
\subsection{Spin-symmetric form factor : leading order}
\begin{figure}
\includegraphics[height=5cm,angle=0]{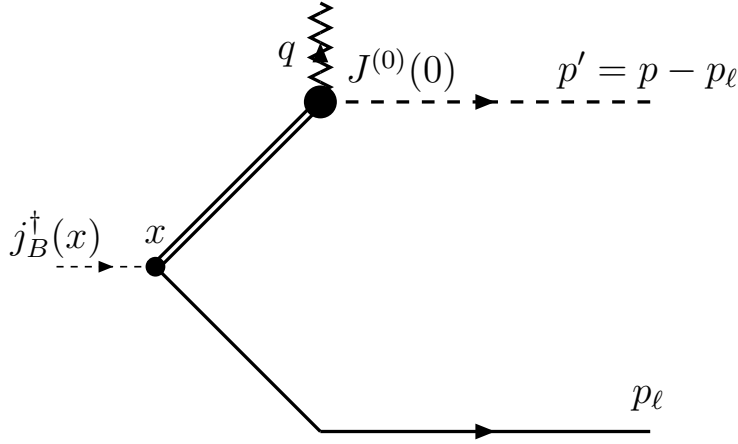}
\caption{\label{Pis}Diagram for the leading order spin-symmetric form factor.}
\end{figure}

Spin-symmetric contributions come from the operators which preserve the
spin-symmetric form factor relations \cite{Bauer:2002aj}.
After the field redefinition of 
$\xi_n\to Y^\dagger \xi_n$, $A_n\to Y^\dagger A_n Y$ where $Y$ is the
soft Wilson line, and scaling down to the ${\rm SCET_{II}}$,
their matrix elements are parameterized as \cite{Beneke:2003pa}
\begin{equation}
\langle\pi(p) |{\bar\xi}_nWY_sh_v(0)|B(m_Bv)\rangle=\nbar\cdot p~\zeta_{sym}~.
\end{equation}
The leading order diagram is shown in Fig.\ \ref{Pis}.
Sum rules for $\zeta_{sym}$ can be obtained via the correlation function
$\Pi_{sym}$:
\begin{equation}
\Pi_{sym}=i\int d^4x~e^{-ip_B\cdot x}\langle\pi(p)|T\left[J^{(0)}(0)j_B^\dagger(x)
\right]|0\rangle~. 
\end{equation}
Using the heavy quark propagator
\begin{equation}
\overbrace{h_v(0)\hbar_v(x)}=\int\frac{d^4k}{(2\pi)^4}
~e^{ik\cdot x}~\frac{i}{v\cdot k+i\epsilon}~\frac{1+\vslash}{2}~,
\end{equation}
we have
\begin{eqnarray}
\Pi_{sym}&=&i\int d^4x~\int\frac{d^4x}{(2\pi)^4}~e^{-i(p_B-m_bv-k)\cdot x}
~\frac{i}{v\cdot k+i\epsilon}\nonumber\\
&&
\times\langle\pi(p)|{\bar\xi}_n(0)\gamma^\mu\left(\frac{1+\vslash}{2}\right)
i\gamma_5 q_s(x)|0\rangle~.
\end{eqnarray}
By expanding $\gamma^\mu$ and $\vslash$ into their light-cone components,
\begin{equation}
{\bar\xi}_n\gamma^\mu\left(\frac{1+\vslash}{2}\right)i\gamma_5 q_s
=
\frac{1}{2}~n^\mu{\bar\xi}_n~\frac{\nbarslash}{2}~i\gamma_5~q_s
+\frac{1}{2}~n^\mu{\bar\xi}_n i\gamma_5~q_s
+\frac{1}{2}~{\bar\xi}_n\gamma_\perp^\mu i\gamma_5~q_s
+\frac{1}{2}~{\bar\xi}_n\gamma_\perp^\mu~\frac{\nbarslash}{2}~i\gamma_5~q_s~.
\end{equation}
Here the terms containing $\gamma_\perp^\mu$ do not contribute, since in the
phenomenological sector the matrix elements of the currents are proportional to
$\sim p_B^\mu$ or $\sim p^\mu$ which has no perpendicular components.
In the full theory, the matrix elements of nonlocal operators between vacuum and
meson are described by the DAs.
Up to the twist 3,
\begin{eqnarray}
\langle\pi(p)|\ubar(0)i\gamma^\mu\gamma_5 d(x)|0\rangle&=&
f_\pi p^\mu\int_0^1 du~e^{i\ubar p\cdot x}\phi_\pi(u)~,\nonumber\\
\langle\pi(p)|\ubar(0)i\gamma_5 d(x)|0\rangle&=&
f_\pi\mu_\pi\int_0^1 du~e^{i\ubar p\cdot x}\phi_p(u)~,\nonumber\\
\langle\pi(p)|\ubar(0)\sigma^{\mu\nu}\gamma_5 d(x)|0\rangle&=&
i\frac{f_\pi\mu_\pi}{6}(p^\mu x^\nu-p^\nu x^\mu)\int_0^1 du~e^{i\ubar p\cdot x}
\phi_\sigma(u)~,
\end{eqnarray}
where $\mu_\pi=m_\pi^2/(m_u+m_d)$.
\par
One delicate point at this stage is how to describe the final state pion.
In the literature, the energetic pion is constructed by two collinear quark fields
in the effective theory.
But this picture does not fully appreciate the soft contributions where the
constituents' momenta are asymmetric.
For the leading order $\zeta_{sym}$, we try to set the relevant matrix element as
\begin{equation}
\langle\pi(p)|{\bar\xi}_n(0)\gamma^\mu\left(\frac{1+\vslash}{2}\right)
i\gamma_5 q_s(x)|0\rangle
=
\frac{n^\mu f_\pi}{4}\int_0^1 du~e^{i\ubar p\cdot x}\left[
\nbar\cdot p~\phi_\pi(u)+2\mu_\pi~\phi_p(u)\right]~,
\label{softDA}
\end{equation}
and consequently,
\begin{equation}
\Pi_{sym}=-\frac{n^\mu f_\pi}{2}\int_0^1 du~
\frac{\eta\phi_\pi(u)+2\mu_\pi\phi_p(u)}{\omega+u\eta-2m_b+i\epsilon}~.
\end{equation}
If the collinear quark $\xi_n$ and the soft quark $q_s$ scale like 
\begin{equation}
\xi_n\sim Q(\lambda^2,1,\lambda)~,~~~{\rm and}~~~
q_s\sim Q(\lambda,\lambda,\lambda)~,
\end{equation}
then the combined momentum has a large virtuality $Q^2\lambda\sim Q\Lambda$.
This is not desirable to form a pion in the final state.
To reduce the large virtuality, it is assumed that the soft quark $q_s$ 
participating in the final pion scales as \cite{Hill:2002vw}
\begin{equation}
q_s\sim (\lambda^2, \lambda, \lambda)~.
\end{equation}
\par
The imaginary part of $\Pi_{sym}$ is proportional to a delta function, which
restricts the range of $u$ after integrating over $s$ in the Borel improved sum
rule  Eq.\ (\ref{BT});
\begin{equation}
u_0<u<1~,~~~1-u_0\equiv\frac{s_0-m_b+\Lmdbar}{m_B}\approx\frac{2\Lmdbar}{m_B}~,
\label{u0}
\end{equation}
where we used the fact that $\omega$ fluctuates around $m_b$ with the amount of 
$\sim\Lmdbar$ and thus its maximum value $s_0$ is roughly $\approx m_b+\Lmdbar$.
The constraint of (\ref{u0}) ensures the collinearity of a parton from the weak
vertex, ${\bar\xi}_n(0)$, and the softness of the spectator quark $q_s(x)$ 
in (\ref{softDA}) \cite{Bagan:1997bp,Ball:2003bf}.
\par
The final result for $\zeta_{sym}$ is
\begin{equation}
\frac{m_B}{m_b}~f_B\eta~\zeta_{sym}
=
\frac{f_\pi}{2}\int_{u_0}^1 du[\eta\phi_\pi(u)+2\mu_\pi\phi_p(u)]
~e^{(2\Lmdbar-\ubar\eta)/T}~,
\end{equation}
where $T$ is the Borel parameter.
In the so called local duality limit where $T\to\infty$,
\begin{equation}
\frac{f_B}{f_\pi}m_b^2~\zeta_{sym} = -\left(\frac{m_b}{m_B}\right)^3~
 \left(\frac{s_0-m_b+\Lmdbar}{2}\right)^2~\phi_\pi'(1)~.
\label{zetas}
\end{equation}
This is exactly the full theory LCSR result \cite{Bagan:1997bp,Ball:2003bf}.
To see this, note that $\omega_0$ in \cite{Bagan:1997bp} is given by
\begin{equation}
\omega_0\equiv v\cdot k|_{\rm max} =\frac{1}{2}(s_0-m_b+\Lmdbar)\approx\Lmdbar~.
\end{equation}
\par
It is quite interesting that Eq.\ (\ref{zetas}) is also consistent with the result of
\cite{DeFazio:2005dx}.
In \cite{DeFazio:2005dx}, the soft form factor $\xi_\pi$ is given by
\begin{equation}
\xi_\pi=\frac{m_B\omega_M}{f_\pi(\nbar\cdot p)}~
\left(1-e^{\omega_s/\omega_M}\right) f_B~\phi_-^B(0)~.
\end{equation}
Then the Wandzura-Wilczek (WW) approximation,
\begin{equation}
\phi_-^B(0)\simeq\int_0^\infty d\omega ~\frac{\phi_+^B(\omega)}{\omega}
\equiv\frac{1}{\lambda_B}~,
\end{equation}
and the sum rule results
\begin{equation}
\nbar\cdot p\omega_M\left(1-e^{\omega_s/\omega_M}\right)\simeq 4\pi^2 f_\pi^2~,
\end{equation}
are used for their analysis.
But if we combine the sum rule result for $1/\lambda_B$ \cite{Ball:2003fq,Ball:2003bf},
\begin{equation}
\frac{1}{\lambda_B}=\frac{3}{2\pi^2}~\omega_0^2~\frac{1}{f_B^2 m_b}~,
\label{lambdaB}
\end{equation}
we arrive at
\begin{equation}
\frac{f_B}{f_\pi}(\nbar\cdot p~ m_b)\xi_\pi=\omega_0^2\cdot 6~.
\end{equation}
This is nothing but the result of (\ref{zetas}) when
$\phi_\pi(u)=\phi_\pi^{\rm asy}(u)\equiv 6u\ubar$.
\subsection{Spin-symmetric form factor : NLO}
\begin{figure}
\begin{tabular}{cc}
\includegraphics[height=5cm,angle=0]{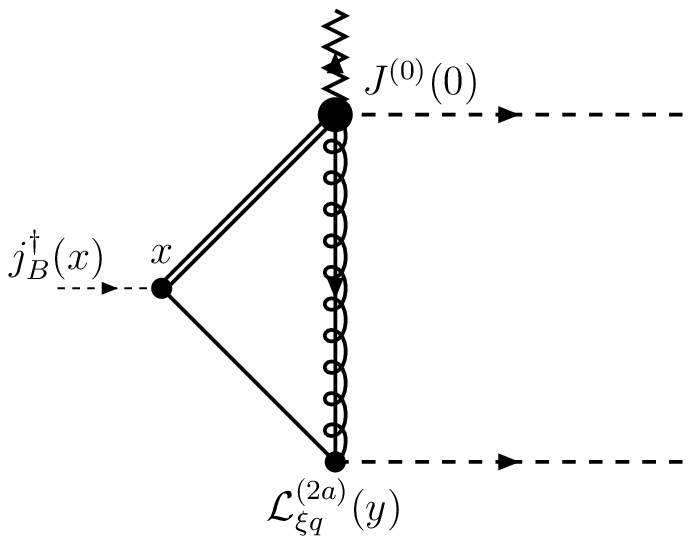}$~~~~~~~~~~~~~~~~~~~~$ &
\includegraphics[height=5cm,angle=0]{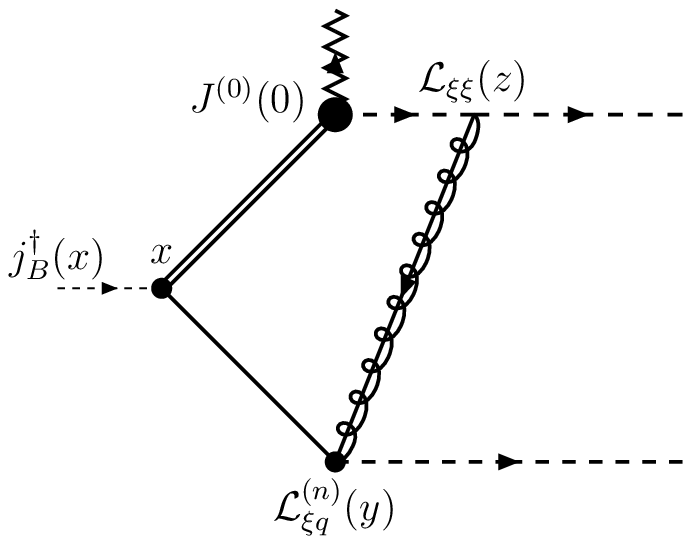}\\
(a) & (b)
\end{tabular}
\caption{\label{Pisym}Diagrams for the NLO spin-symmetric form factors.}
\end{figure}

At NLO of $\alpha_s$, $\zeta_{sym}$ includes spectator interactions with collinear
gluon exchanges.
The spectator quark hit by the collinear gluon becomes collinear quark at this order,
so the momentum configuration of the partons can be symmetric.
Figure \ref{Pisym} shows some of them.
As an illustration, we construct the correlation functions for these diagrams as
\begin{eqnarray}
\Pi_{sym}^{(0,2a)}&=&i\int d^4x~e^{-ip_B\cdot x}\langle\pi(p)|T\left[
J^{(0)}(0)j_B^\dagger(x),\int d^4y~\calL_{\xi q}^{(2a)}(y)\right]|0\rangle~,
\nonumber\\
\Pi_{sym}^{(n)}&=&i\int d^4x~e^{-ip_B\cdot x}\langle\pi(p)|T\left[
J^{(0)}(0)j_B^\dagger(x),
\int d^4z~\calL_{\xi\xi}^{(0)}(z),
\int d^4y~\calL_{\xi q}^{(n)}(y)\right]0\rangle~.
\label{symNLO}
\end{eqnarray}
Note the presence of the vertex $\calL_{\xi q}$ which converts the usoft spectator 
into a collinear quark.
With the dimensional regularization where the dimension $d=4-2\epsilon$, we have
\begin{eqnarray}
\frac{1}{\pi}{\rm Im}\left[\Pi_{sym}^{(0,2a)}\right]&=&
\frac{g^2C_F}{16\pi^2}~f_\pi n^\mu\int_0^1 du ~\frac{\phi_\pi(u)}{\ubar}\nonumber\\
&&\times\left[
-\frac{1}{\epsilon}~\frac{r}{\ubar-r}
+\frac{2r}{\ubar-r}~\ln\left(\frac{r\eta}{\mu}\right)
-\frac{\ubar}{\ubar-r}~\ln\left(\frac{\ubar}{\ubar-r}\right)\right]~,
\nonumber\\
\frac{1}{\pi}{\rm Im}\left[\Pi_{sym}^{(1)}\right]&=&
\frac{g^2C_F}{16\pi^2}~f_\pi n^\mu\int_0^1 du ~\frac{\phi_\pi(u)}{\ubar}\nonumber\\
&&\times u\left
[\frac{1}{\epsilon}~\frac{r}{\ubar-r}
-\frac{2r}{\ubar-r}~\ln\left(\frac{r\eta}{\mu}\right)
+\frac{\ubar}{\ubar-r}~\ln\left(\frac{\ubar}{\ubar-r}\right)
\right]~,\nonumber\\
\frac{1}{\pi}{\rm Im}\left[\Pi_{sym}^{(2a)}\right]&=&
\frac{g^2C_F}{16\pi^2}~f_\pi n^\mu\int_0^1 du ~\frac{\phi_\pi(u)}{\ubar}\nonumber\\
&&\times\left[
-\frac{1}{\epsilon}~\frac{r(1-r)}{\ubar-r}
+\frac{2r(1-r)}{\ubar-r}~\ln\left(\frac{r\eta}{\mu}\right)-r
-\frac{u\ubar}{\ubar-r}~\ln\left(\frac{\ubar}{\ubar-r}\right)
\right]~,\nonumber\\
\frac{1}{\pi}{\rm Im}\left[\Pi_{sym}^{(2b)}\right]&=&
\frac{g^2C_F}{16\pi^2}~f_\pi n^\mu\int_0^1 du ~\frac{\phi_\pi(u)}{\ubar}
\left[-\frac{1}{\epsilon}~\frac{ur^2}{(\ubar-r)^2}
+\frac{2ur^2}{(\ubar-r)^2}~\ln\left(\frac{r\eta}{\mu}\right)\right.\nonumber\\
&&
\left.
-\frac{ru\ubar}{(\ubar-r)^2}
+\frac{\ubar}{(\ubar-r)^2}(-2ru+\ubar+\ubar^2)~\ln\left(\frac{\ubar}{\ubar-r}\right)
\right]~,
\end{eqnarray}
where $r\equiv(\omega+\eta-2m_b)/\eta$.
Since $\omega$ describes a small fluctuation of order $\calO(\Lmdbar)$ around
$m_ b$, $r=\calO(\Lmdbar/m_B)\ll 1$ when $\eta=\calO(m_B)$.
\par
Summing up these terms gives the finite part as
\begin{eqnarray}
\frac{1}{\pi}{\rm Im}\Pi_{sym}^{\rm NLO}&\equiv&\frac{1}{\pi}{\rm Im}\left[
\Pi_{sym}^{(0,2a)}+\Pi_{sym}^{(1)}+\Pi_{sym}^{(2a)}+\Pi_{sym}^{(2b)}\right]
\nonumber\\
&=&
\frac{g^2C_F}{16\pi^2}~f_\pi n^\mu\int_0^1 du ~\frac{\phi_\pi(u)}{\ubar}\left[
-r
+\frac{2r}{\ubar-r}\left\{\ubar+(1-r)+\frac{ur}{\ubar-r}\right\}
\ln\left(\frac{r\eta}{\mu}\right)\right.\nonumber\\
&&
\left.
-\frac{ru\ubar}{(\ubar-r)^2}
+\frac{\ubar}{(\ubar-r)^2}\left\{\ubar^2+r(1-2u)\right\}
\ln\left(\frac{\ubar}{\ubar-r}\right)\right]~.
\end{eqnarray}
In the limit of $r\to 0$,
\begin{equation}
\frac{1}{\pi}{\rm Im}\Pi_{sym}^{\rm NLO}=
\frac{g^2C_F}{16\pi^2}~f_\pi n^\mu\int_0^1 du ~\phi_\pi(u)\left[
2r\left(\frac{1}{\ubar}+\frac{1}{\ubar^2}\right)
\ln\left(\frac{r\eta}{\mu}\right)+\frac{r}{\ubar}\left(1-\frac{1}{\ubar}\right)\right]~.
\end{equation}
The light-cone sum rule, Eq.\ (\ref{BT}), is now established as
\begin{equation}
\frac{f_B}{f_\pi}m_b^2 \zeta_{sym}^{\rm NLO}
=\frac{g^2C_F}{4\pi^2}\Lmdbar^2\int_0^1 du~\phi_\pi(u)\left[
\left(\frac{1}{\ubar}+\frac{1}{\ubar^2}\right)
 \ln\left(\frac{2\Lmdbar}{\mu}\right)-\frac{1}{\ubar^2}\right]~,
\end{equation}
when $T\to\infty$.
This result must be compared with Eq.\ (14) of \cite{Bagan:1997bp}.
Terms which are not proportional to $\phi_\pi'(1)$ are successfully reproduced.
\subsection{Spin non-symmetric form factor}
\par
\begin{figure}
\includegraphics[height=5cm,angle=0]{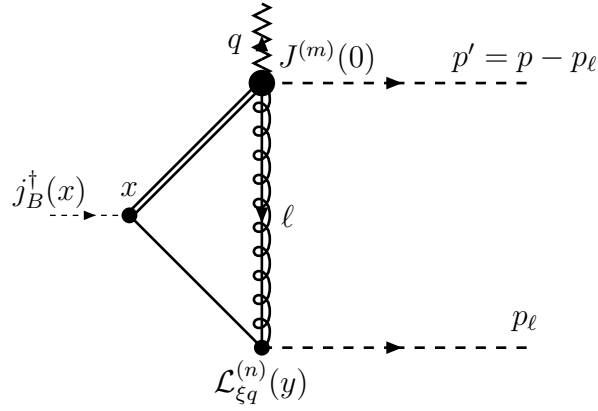}
\caption{\label{PiF}Diagram for the spin non-symmetric form factors where
$m=1a,~1b$.}
\end{figure}

The matrix elements of power suppressed currents $J^{(1a,1b)}$ are proportional
to the spin non-symmetric form factors.
Nonzero contributions involve a collinear gluon exchange between the weak vertex
and the usoft spectator which becomes collinear after the interaction.
Thus the correlation function which gives sum rules for the factorizable form
factors can be written as ($m=1a,~1b$)
\begin{equation}
\Pi_{ns}^{(m,n)}=i\int d^4x~e^{-ip_B\cdot x}\langle\pi(p)|T\left[
J^{(m)}(0)j_B^\dagger(x),\int d^4y~\calL_{\xi q}^{(n)}(y)\right]|0\rangle~.
\label{Pimn}
\end{equation}
Figure \ref{PiF} shows the $\Pi_F^{(m,n)}$ and kinematics.
Using the Feynman rules for $J^{(m)}$ and $\calL_{\xi q}^{(n)}$ from 
\cite{Pirjol:2002km}, we have
\begin{eqnarray}
\frac{1}{\pi}{\rm Im}\left[\Pi_{ns}^{(1a,1)}\right]&=&
\frac{g^2 C_F}{16\pi^2}f_\pi\nbar^\mu\int_0^1 du~\frac{\phi_\pi(u)}{\ubar}
 \ln\frac{1}{1-r}~,
\nonumber\\
\frac{1}{\pi}{\rm Im}\left[\Pi_{ns}^{(1a,2a)}\right]&=&0~,
\nonumber\\
\frac{1}{\pi}{\rm Im}\left[\Pi_{ns}^{(1a,2b)}\right]&=&
\frac{g^2 C_F}{16\pi^2}f_\pi\nbar^\mu\int_0^1 du~\frac{\phi_\pi(u)}{u\ubar}
\left(\ln\frac{1}{1-r}-\ubar\ln\frac{\ubar}{\ubar-r}\right)~,
\nonumber\\
\frac{1}{\pi}{\rm Im}\left[\Pi_{ns}^{(1b,1)}\right]&=&
\frac{g^2 C_F}{16\pi^2}f_\pi n^\mu\int_0^1 du~\frac{\phi_\pi(u)}{\ubar}
~\frac{r\eta}{m_b}~,
\nonumber\\
\frac{1}{\pi}{\rm Im}\left[\Pi_{ns}^{(1b,2a)}\right]&=&
\frac{g^2 C_F}{16\pi^2}f_\pi n^\mu\int_0^1 du~\frac{\phi_\pi(u)}{\ubar}
~\frac{\eta}{m_b}\left(r-\ubar\ln\frac{\ubar}{\ubar-r}\right)~,
\nonumber\\
\frac{1}{\pi}{\rm Im}\left[\Pi_{ns}^{(1b,2b)}\right]&=&
-\frac{1}{\pi}{\rm Im}\left[\Pi_{ns}^{(1b,2a)}\right]~.
\label{results}
\end{eqnarray}
It is easy to check that the above results reproduce the previous calculations
for the factorizable form factors by full theory LCSR and QCDF.
To see this, first consider the sum rule for $J^{(1b)}$.
Since
${\rm Im}\left[\Pi_{ns}^{(1b,2b)}\right]=-{\rm Im}\left[\Pi_{ns}^{(1b,2a)}\right]$,
we have
\begin{eqnarray}
\lefteqn{
\frac{F_B}{\sqrt{m_B}}~e^{-(2m_B-\eta)/T}\langle\pi|J^{(1b)}|B\rangle}
\nonumber\\
&=&
\frac{1}{\pi}\int_{m_b-\Lmdbar}^{s_0}{\rm Im}\left[\Pi_{ns}^{(1b,1)}(s,\eta)\right]
e^{-s/T}~ds\nonumber\\
&=&
\frac{g^2C_F}{16\pi^2}f_\pi n^\mu\int_{m_b-\Lmdbar}^{s_0}~ds~e^{-s/T}
\int_0^1~du~\frac{\phi_\pi(u)}{\ubar}~\left(\frac{s+\eta-2m_b}{m_b}\right)~.
\end{eqnarray}
In case of $T\to\infty$ and $\eta=\nbar\cdot p=2E_\pi=m_B~(q^2=0)$,
the right-hand-side of sum rule becomes
\begin{equation}
({\rm R.H.S})=\frac{g^2C_F}{16\pi^2}f_\pi n^\mu\langle\ubar^{-1}\rangle_\pi
\frac{1}{2m_b}(s_0-m_b+\Lmdbar)^2~,
\label{RHS}
\end{equation}
where
\begin{equation}
\langle\ubar^{-1}\rangle_\pi=\int_0^1~du~\frac{\phi_\pi(u)}{\ubar}~.
\end{equation}
On the other hand, the left-hand-side is
\begin{equation}
({\rm L.H.S})=\frac{F_B}{\sqrt{m_B}}\langle\pi|\left(-\frac{1}{m_b}\right)
\bar{\xi}_n\gamma^\mu~\frac{\nslash}{2}~i\Dslash_c^\perp~Wh_v|B\rangle~.
\label{LHS}
\end{equation}
The matrix element is proportional to the factorizable form factor
$\Delta F_\pi$ in QCDF \cite{Beneke:2000wa,Beneke:2003pa,DeFazio:2005dx} :
\begin{equation}
-\langle\pi|\bar{\xi}_n i\Dslash_c^\perp~Wh_v|B\rangle
=\frac{m_B^2}{2}~\frac{g^2C_F}{16\pi^2}~\Delta F_\pi~.
\label{DeltaF}
\end{equation}
After contracting $\nbar_\mu$ on both sides, combining Eqs.\ (\ref{RHS}),
(\ref{LHS}), and (\ref{DeltaF}) provides the sum rule for $\Delta F_\pi$:
\begin{equation}
\Delta F_\pi=\frac{m_b}{m_B^3 f_B}~f_\pi\langle\ubar^{-1}\rangle_\pi
(s_0-m_b+\Lmdbar)^2~.
\label{SR1}
\end{equation}
Now that the well-known QCDF result for $\Delta F_\pi$ is
\begin{equation}
\Delta F_\pi
=\frac{8\pi^2f_Bf_\pi}{3m_B}\lambda_B^{-1}\langle\ubar^{-1}\rangle_\pi~,
\end{equation}
Eq.\ (\ref{SR1}) is consistent with QCDF provided that
\begin{equation}
\lambda_B^{-1}=\frac{3}{2\pi^2}~\frac{m_b}{m_B^2 f_B^2}~
\left(\frac{s_0-m_b+\Lmdbar}{2}\right)^2~.
\label{lambdaB2}
\end{equation}
This is indeed the case as already mentioned in Eq.\ (\ref{lambdaB}).
\par
Sum rules for $J^{(1a)}$ also give the same result for $\Delta F_\pi$.
Since $r$ is a small quantity in Eq.\ (\ref{results}),
\begin{equation}
{\rm Im}\left[\Pi_{ns}^{(1a,1)}\right]\sim\ln\frac{1}{1-r}
=r+\calO(r^2)\approx{\rm Im}\left[\Pi_{ns}^{(1b,1)}\right]~.
\end{equation}
As for ${\rm Im}\left[\Pi_{ns}^{(1a,2b)}\right]$, we can estimate its size after
integrating over $u$ with the asymptotic form of $\phi_\pi(u)$,
$\phi_\pi^{\rm asy}(u)=6u\ubar$:
\begin{equation}
{\rm Im}\left[\Pi_{ns}^{(1a,2b)}\right]\sim
\int_0^1~du~\frac{\phi_\pi^{\rm asy}}{u\ubar}\left(
\ln\frac{1}{1-r}-\ubar\ln\frac{\ubar}{\ubar-r}\right)
=\calO(r^2)~.
\end{equation}

\section{Discussions and conclusions}

In the present analysis, $\omega=2v\cdot q$ describes a small fluctuation of order
$\calO(\Lmdbar)$ around $m_ b$;
\begin{equation}
m_b-\Lmdbar\le\omega\le m_b+\Lmdbar~,
\end{equation}
so $s_0\approx m_b+\Lmdbar$.
This parameterization is consistent with $\omega_0$ in \cite{Bagan:1997bp} when
$\omega_0=\Lmdbar$.
For a numerical analysis one usually treats $s_0$ (or $\omega_0$) as a free 
parameter with some reasonable constraints, but the numerics will not be 
considered here.
\par
As shown in the previous section, the sum rule result from \cite{DeFazio:2005dx}
for the ''soft form factor'' is coincident with that of this work or the previous
conventional LCSR.
The reason is that when the final state pion is described by the interpolating
current, only terms like
\begin{equation}
J_\pi(x)\sim {\bar\xi}_{\rm hc}W_{\rm hc}(x)\nbarslash\gamma_5Y_s^\dagger
q_s(x)+h.c.
\label{Jpi}
\end{equation}
are contributing to the ''soft form factor'' (or, the leading spin-symmetric
form factor) $\xi_\pi$.
Here the soft field $q_s(x)$ should exist to define the $B$ meson distribution
amplitude together with the heavy quark field $h_v$ from the weak current.
In this picture, the presence of both ${\bar\xi}$ and $q_s$ at the point $x$ is
{\em a priori}, without any dynamical explanations.
The original problem of how to form an energetic pion in the final state with
one soft and one collinear quark remains in principle unresolved.
Descriptions of Eqs.\ (\ref{softDA}) and (\ref{Jpi}) are {\em equivalent} in the
sense that the pion is depicted by highly momentum-asymmetric configuration with
the collinear field $\bar\xi$ and the soft field $q_s$.
The potential problem of large nonlocality in Eq.\ (\ref{softDA}) is alleviated by
requiring an accidental smallness of the plus component of $q_s$,
$q_s\sim (\lambda^2, \cdots)$ \cite{Hill:2002vw}.
\par
We classify the collinear gluon exchange diagrams for the spin-symmetric currents
as the NLO correction to the spin-symmetric form factor.
Note that the operators appearing in Eq.\ (\ref{symNLO}) are similar to the
''non-factorizable'' operators of \cite{Bauer:2002aj}.
But the non-factorizable operators contain the soft gluons in a nontrivial way,
which makes it difficult to factorize their matrix elements.
In general, the ''non-factorizable'' soft gluon effects can include the three-particle
distribution amplitudes of the form \cite{Ball:1998je}
\begin{equation}
\langle 0|\ubar(z)\gamma_\mu\gamma_5 g G_{\alpha\beta}~d(-z)|\pi\rangle~,~~~
{\rm or}~~~
\langle 0|\ubar(z)\gamma_\mu ig {\tilde G}_{\alpha\beta}~d(-z)|\pi\rangle~,
\end{equation}
where $G_{\alpha\beta}({\tilde G}_{\alpha\beta})$ is the gluon field strength.
The importance of three-particle DAs was already pointed out in
\cite{Beneke:2003pa}.
They introduce additional four distribution functions of twist 4.
Thus it would be quite interesting to investigate the three-particle DAs in the
full theory and in SCET and compare the results.
The analysis will check if the operator set of \cite{Bauer:2002aj} is complete
at this accuracy.
But as for the predictive power, there will be few improvements because
three-particle DA analysis trades single nonperturbative parameter
$\zeta^{B\pi}$ for four more DAs.
\par
It must be emphasized that the NLO spin-symmetric contributions contain potential
end-point singularity terms $\sim 1/\ubar^2$.
In the full theory, these terms are combined with terms $\sim \phi_\pi'(1)/\ubar$
to regularize the end-point divergence \cite{Bagan:1997bp,Ball:2003bf}.
We expect a similar situation in the effective theory.
Terms like $\sim \phi_\pi'(1)/\ubar$ would appear for the diagrams of 
spectator interactions with soft gluons.
Since the Feynman rules of the soft sector are the same as those in the full QCD,
momentum-asymmetric features are exactly reproduced in the effective theory.
And also, this should happen to satisfy the evolution equation of $\phi_\pi'(1,\mu)$
in the full theory \cite{Bagan:1997bp}.
\par
Spin-nonsymmetric contributions are exactly the ''factorizable'' part of
\cite{Bauer:2002aj}, or  $\Delta F_\pi$ in \cite{Beneke:2000wa,Beneke:2003pa}.
In SCET, both $B$ and $\pi$ are described by the corresponding DAs in a 
convoluted manner :
\begin{equation}
\Delta F_\pi\sim\phi_B\otimes J\otimes\phi_\pi~,
\end{equation}
where $J$ is the jet function.
But in LCSR, the initial $B$ meson is described by an interpolating current.
Thus a comparison between LCSR and SCET will provide some relation for $\phi_B$,
or the moment of $\phi_B$, more exactly.
Actually, the relation is no other than Eq.\ (\ref{lambdaB2}).
It is remarkable that $\lambda_B^{-1}$ from this work is coincident with that
from LCSR in $B\to\gamma e\nu$ \cite{Ball:2003fq}.
On the other hand, a new approach of \cite{DeFazio:2005dx} takes the pion to be
described by an interpolating current.
Consequently, a comparison between \cite{DeFazio:2005dx} and SCET gives a 
relation for the moment of $\phi_\pi~$, $\langle u^{-1}\rangle_\pi\simeq 3$.
In this sense, the present work and \cite{DeFazio:2005dx} is complementary.
\par
For the NLO spin-symmetric or spin-nonsymmetric contributions where the collinear
gluon is exchanged, the main motivations of new sum rule of \cite{DeFazio:2005dx}
become weak since the final quarks are all collinear.
Further, what we are mainly concerned about heavy-to-light decay is {\em how to}
form an energetic light meson in the final state.
In the new approach, the final state meson is described by a local interpolating
current, leaving other information in the $B$ meson DA.
Another merit of the conventional LCSR over the new one is that the variety
of $B$ decays can be easily encapsulated by the final-state mesons' DAs.
\par
As for the numerical size of each contributions, present analysis provides
nothing new.
Relative size of each component of the form factor is still very disputable
\cite{Chay}, and more efforts should be made in this direction.
\par
In summary, heavy-to-light decay form factor is reexamined in the framework of
LCSR in SCET.
In this work, unlike a recent approach in \cite{DeFazio:2005dx}, conventional
LCSR is naturally extended for an effective theory.
Establishments of the sum rules in the effective theory is useful in that
different kinematic configurations are separated from the beginning at the
operator level.
After assembling all the pieces, we successfully reproduce the full theory
results for the form factor.
There still remain complicated and delicate problems of how to describe the
Feynman mechanism clearly with operators in the effective theory, or how large
each contributions is.
\begin{acknowledgments}
The author thanks Junegone Chay for helpful discussions.
\end{acknowledgments}

\end{document}